\documentclass[namedreferences]{SolarPhysicsMod}
\usepackage[optionalrh]{spr-sola-addons} 
\usepackage{solaheader}
\newcommand{\solareceiveddate}{13 November 2009}
\newcommand{\solaaccepteddate}{24 June 2010}
\newcommand{\solaonlinedate}{28 August 2010}
\newcommand{\doi}{10.1007/s11207-010-9611-7}
\usepackage{graphicx}                    
\usepackage{color}                       
\usepackage{url}                         

\begin{document}

\begin{article}

\begin{opening}

\title{Determining Absorption, Emissivity Reduction, and Local Suppression Coefficients inside Sunspots}

%
\author{Stathis ~\surname{Ilonidis}\sep
        Junwei ~\surname{Zhao}
       }

%
\runningauthor{S. Ilonidis, J. Zhao} \runningtitle{Absorption,
Emissivity Reduction, and Local Suppression}

%
  \institute{W.W. Hansen Experimental Physics Laboratory, Stanford University,
Stanford, CA 94305-4085, USA \\ email:
\url{ilonidis@sun.stanford.edu}
             }

\begin{abstract}
The power of solar acoustic waves is reduced inside sunspots mainly
due to absorption, emissivity reduction, and local suppression. The
coefficients of these power-reduction mechanisms can be determined
by comparing time-distance cross-covariances obtained from sunspots
and from the quiet Sun. By analyzing $47$ active regions observed by
SOHO/MDI without using signal filters, we have determined the
coefficients of surface absorption, deep absorption, emissivity
reduction, and local suppression. The dissipation in the quiet Sun
is derived as well. All of the cross-covariances are width corrected
to offset the effect of dispersion. We find that absorption is the
dominant mechanism of the power deficit in sunspots for short travel
distances, but gradually drops to zero at travel distances longer
than about $6^\circ$. The absorption in sunspot interiors is also
significant. The emissivity-reduction coefficient ranges from about
$0.44$ to $1.00$ within the umbra and $0.29$ to $0.72$ in the
sunspot, and accounts for only about $21.5\%$ of the umbra's and
$16.5\%$ of the sunspot's total power reduction. Local suppression
is nearly constant as a function of travel distance with values of
$0.80$ and $0.665$ for umbrae and whole sunspots respectively, and
is the major cause of the power deficit at large travel distances.

\end{abstract}

%

\end{opening}

%
\section{Introduction}
\par Solar oscillations  have lower amplitude in sunspots than in the
quiet Sun \linebreak (Leighton, Noyes, and Simon, 1962; Lites, White, and
Packman, 1982). The observed power reduction is caused by several
mechanisms which, in this paper, are grouped in three categories:
absorption, emissivity reduction, and local suppression, as
categorized by Chou \emph{et al.} (2009c).
\par The first category, absorption, includes three different mechanisms: \textit{\textrm{i}}) acoustic
wave energy is converted into heat (Hollweg, 1988; Lou, 1990;
Sakurai, Goossens, and Hollweg, 1991; Goossens and Poedts, 1992),
\textit{\textrm{ii}}) acoustic waves are converted into modes that
cannot be detected in the photosphere (Spruit and Bogdan, 1992;
Cally and Bogdan, 1993; Cally, Bogdan, and Zweibel, 1994; Crouch and
Cally, 2005; Gordovskyy and Jain, 2008), \textit{\textrm{iii}}) wave
leakage to the outer atmosphere is enhanced due to the modification
of the acoustic cut-off frequency. The second category is emissivity
reduction. The reduction in convection in sunspots leads to reduced
excitation of acoustic waves (Hurlburt and Toomre, 1988; Parchevsky
and Kosovichev, 2007). The last category, local suppression, is a
local change in observed wave amplitude, rather than a change in
energy. Several mechanisms also fall into this category. The
observed amplitude reduction may be caused by: \textit{\textrm{i}})
Wilson depression (Hindman, Jain, and Zweibel, 1997),
\textit{\textrm{ii}}) altered eigenfunctions (Hindman, Jain, and
Zweibel, 1997), \textit{\textrm{iii}}) a greater wave speed, and
\textit{\textrm{iv}}) a change in the line profile (Wachter, Schou,
and Sankarasubramanian, 2006; Rajaguru \emph{et al.}, 2007). For
more details on each mechanism see Hindman, Jain, and Zweibel (1997)
and Chou \emph{et al.} (2009a, 2009b).
\par Braun, Duvall, and LaBonte (1987, 1988), Bogdan \emph{et al.}
(1993), and Chen, Chou, and the TON team (1996) found, using Hankel analysis
and spherical decomposition of the acoustic wavefield, that sunspots
absorb as much as $50\%$ of the incoming acoustic waves. This method
includes decomposition of the \textit{p}-mode oscillations into
inward and outward propagating modes with respect to the sunspot.
However, the effects of absorption and emissivity reduction cannot
be distinguished, and, additionally, local suppression is not
included in such measurements.
\par The first attempt to distinguish and determine the three coefficients of absorption,
emissivity reduction, and local suppression was made by Chou
\emph{et al.} (2009b, 2009c). For this purpose they used the
property that the waves emitted along the wave path between two
points have no correlation with the signal at the starting point.
Their technique makes use of direction and phase-velocity filters,
which successfully allow measurements of the three above coefficients
for specific acoustic wave-travel distances.
\par In this work we use a similar yet new method, which does not make use
of the direction and phase-velocity filters to determine the
coefficients of absorption, emissivity reduction, and local
suppression. Instead, we use many active regions to increase the
signal-to-noise ratio. The measurement procedure, in the absence of
those filters, is more general and does not depend on the particular
characteristics of the filters. More specifically, the absence of
the phase-velocity filter allows the determination of all three
parameters as functions of travel distance from active regions,
while the absence of direction filter does not raise any questions
regarding the width of the filter. This issue is important in some
steps of the measurement procedure such as the correction for the
dispersion of acoustic waves. On the other hand, the use of many
active regions limits our method to an averaged measurement of
absorption, emissivity reduction, and local suppression inside
active regions only as functions of travel distance from the center
of the regions but not as functions of direction.

\section{Method}
\begin{figure}    
   \centerline{\hspace*{-0.0\textwidth}
               \includegraphics[width=0.7\textwidth,clip=]{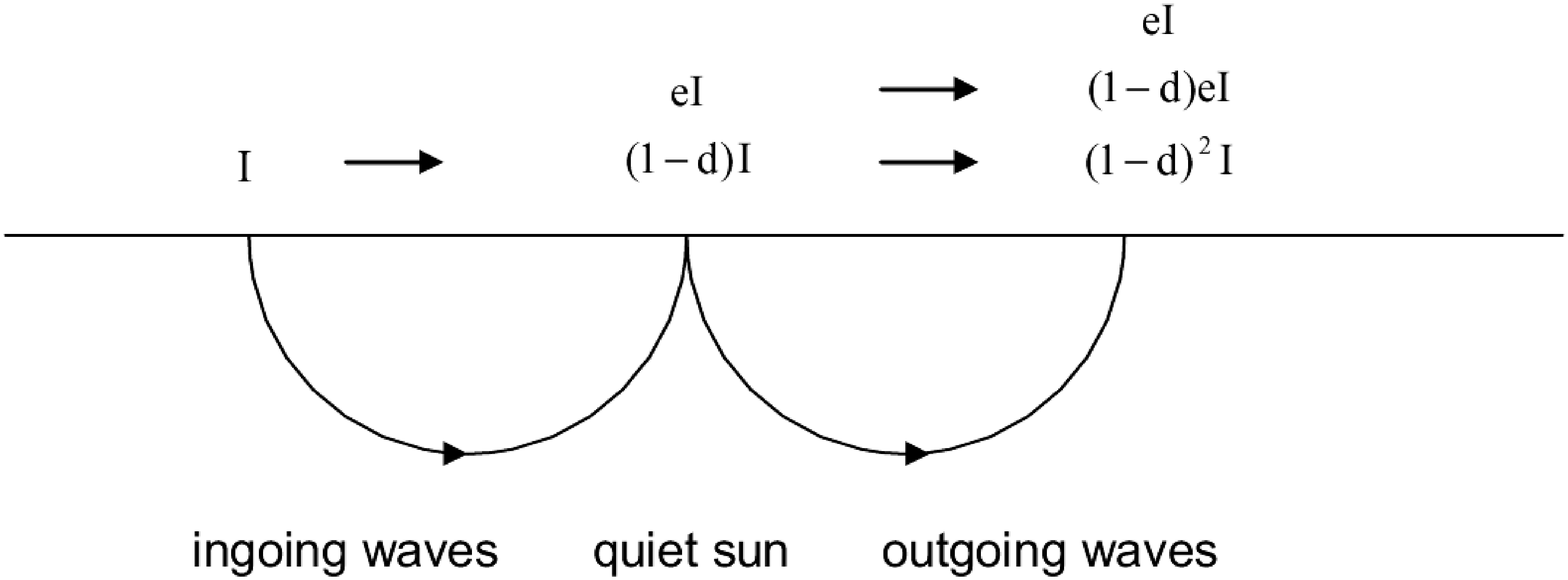}
              }
              \vspace{0.0\textwidth}
   \centerline{\hspace*{-0.0\textwidth}
               \includegraphics[width=0.7\textwidth,clip=]{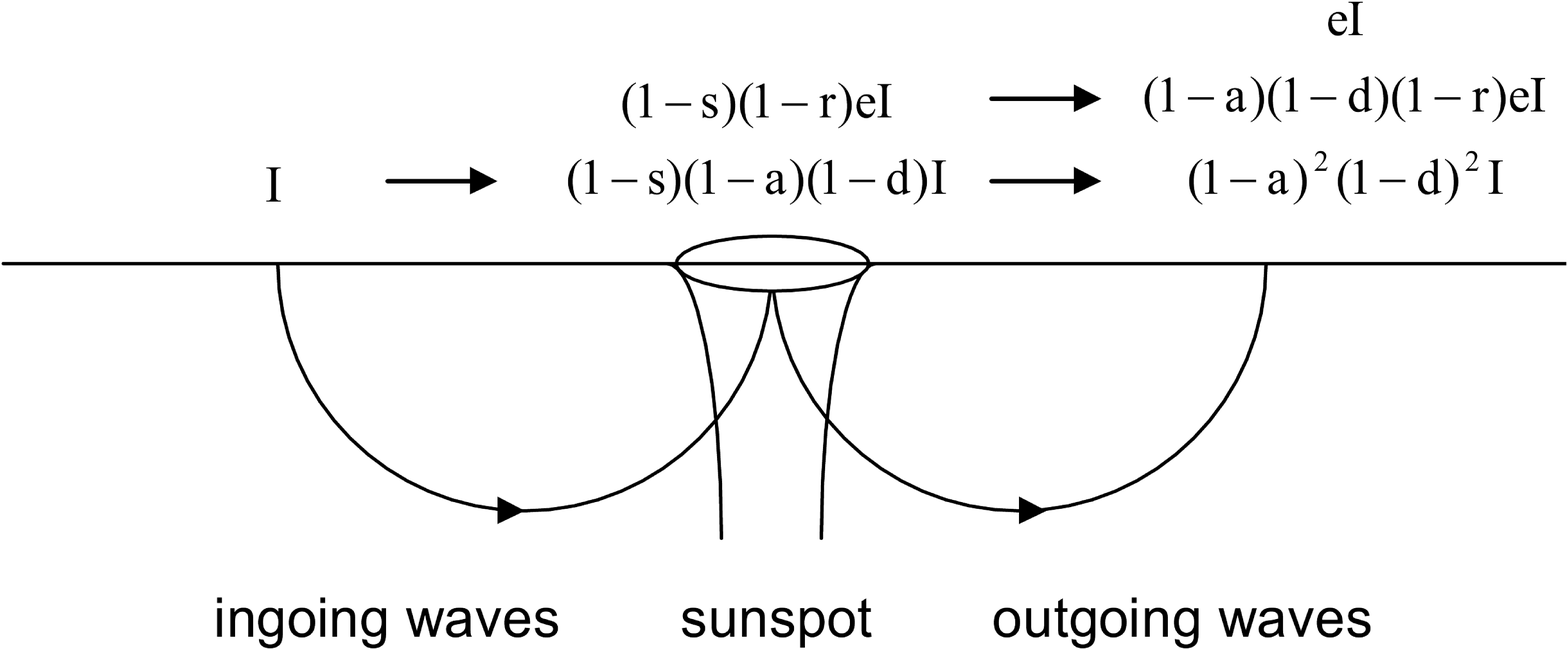}
              }
               \vspace*{0.0\textwidth}
   \centerline{\hspace*{-0.0\textwidth}
               \includegraphics[width=0.7\textwidth,clip=]{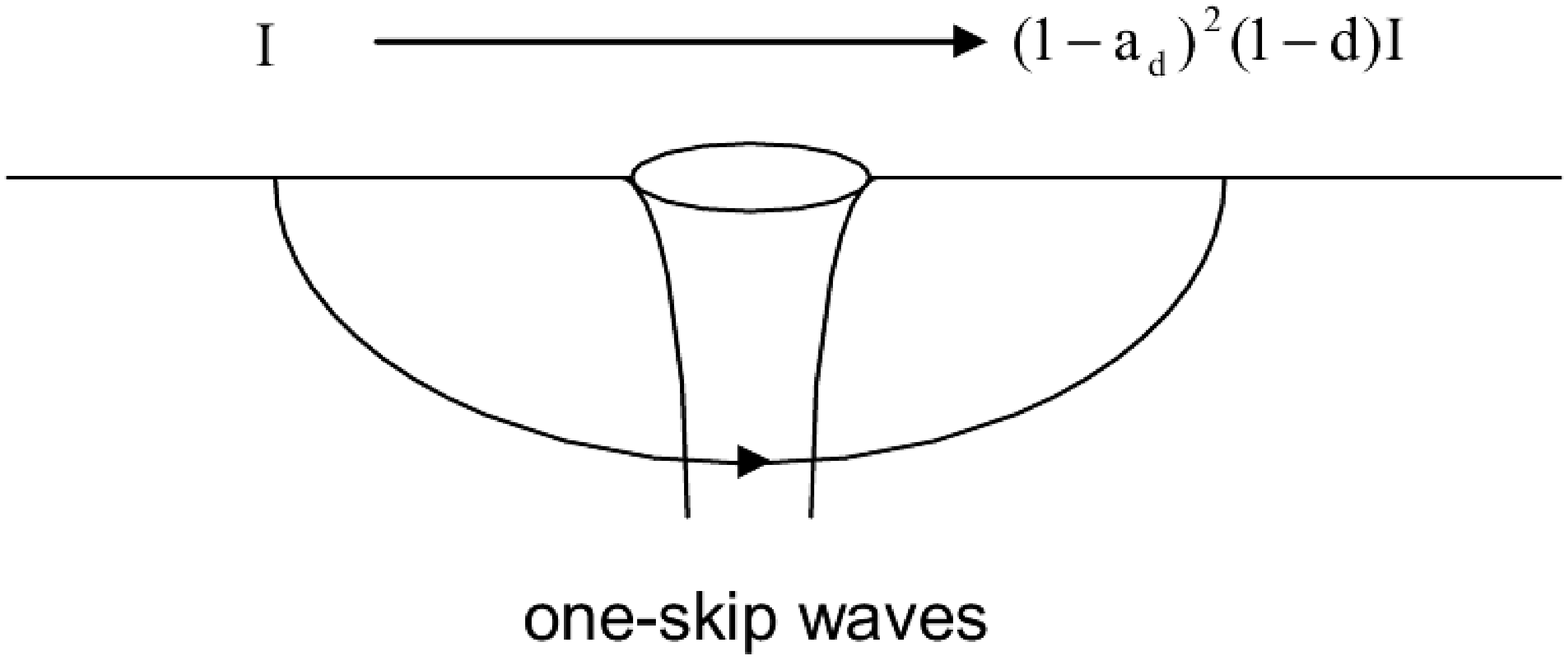}
              }
               \vspace{0.\textwidth}    

              \caption{Energy budget of ingoing, outgoing, and double-skip waves propagating through
              the quiet Sun (top), ingoing, outgoing, and double-skip waves propagating through a sunspot (middle),
              and one-skip waves propagating through a sunspot
              (bottom). Only terms connected by arrows correlate.
                       }
   \label{energy_budget}
   \end{figure}

The energy budget of acoustic waves propagating through the quiet
Sun and a sunspot is illustrated in Figure \ref{energy_budget}. The
energy of a wave packet associated with a particular travel distance
and propagating in a particular direction in the quiet Sun is
constant and is denoted here by $I$. As the waves propagate through
the solar medium in the top diagram of Figure \ref{energy_budget},
the acoustic energy after one skip is reduced by a factor of
$(1-d(r))$ due to dissipation, where $d$ is the dissipation
coefficient in the quiet Sun and $r$ is the one-skip travel
distance. At the same time new waves with energy $e(r)I$ are
generated by turbulent convection where $e(r)$ is the emissivity
coefficient in the quiet Sun. Since the acoustic energy is constant
in the quiet Sun, $e(r)=d(r)$.

\par For the two
signals, $(1-d(r))I$ and $e(r)I$, only the first one correlates with
the signal $I$ because $e(r)I$ is generated near the location of
reflection by a different source. We use the above property to
determine the dissipation of acoustic waves $\mathit{d(r)}$ in the quiet Sun.
If the energy $I$ of the acoustic waves is reduced by a factor of
$(1-d(r))$ after each skip, the energy of the same signal after $n$
skips will be $(1-d(r))^nI$. Using the definition of
cross-covariance between two points as given in Equation (1), we
compute the ratio of amplitude of the cross-covariance at the fifth
skip, $F_{5}$, to that at the first skip $F_{1}$. Here the wave
function $\Psi$ is the square root of the energy. According to Equation
(\ref{correlation}), $F_{5}=(1-d)^{5/2}I$ and $F_{1}=\sqrt{(1-d)}I$.
Equation (\ref{dissipation}) gives the dissipation in terms of
$F_{5}$ and $F_{1}$. The exact computational procedure used to
determine $d(r)$ is described in the second paragraph of Section 3.

\begin{equation}
F_{12}(\tau)=\sum_{t}\Psi_{1}(t)\Psi_{2}(t+\tau) \label{correlation}
\end{equation}

\begin{equation}
d(r)=1-\left(\frac{F_{5}}{F_{1}}\right)^{1/2} \label{dissipation}
\end{equation}

\par For the energy budget of the acoustic waves in the sunspot, we follow the
model suggested by Chou \emph{et al.} (2009c) but incorporate a
minor modification. According to this model the terms $(1-d(r))I$
and $e(r)I$ that were mentioned before in the quiet Sun are reduced by
the factors $(1-a(r))(1-s(r))$ and $(1-s(r))(1-r(r))$ respectively
inside the sunspot, where $a$ is the absorption coefficient, $r$ is
the emissivity reduction coefficient, and $s$ is the local
suppression coefficient. After one more skip, the acoustic waves
propagate outside the sunspot in the region of the quiet Sun. Both
terms are further reduced by a factor of $(1-a(r))(1-d(r))$ and
amplified by a factor of $(1-s(r))$ because local suppression
disappears outside the sunspot. At the location of the second skip,
in addition to the terms described above, new waves with energy
$e(r)I$ are generated by turbulent convection. For more details of
this model see Chou \emph{et al.} (2009c).

\par We consider now a reference point on the surface of the
quiet Sun. Using the definition of cross-covariance as given in
Equation (\ref{correlation}), we compute the amplitude of
cross-covariances of ingoing, outgoing, and two-skip waves with
respect to the reference point as functions of travel distance from
that point. The two-skip waves have a reflection located in the
reference point. The exactly same procedure is applied to the case
of a sunspot where the reference point is always inside the sunspot.
Considering the ratios of amplitude of the cross-covariance for the
ingoing, outgoing and two-skip waves in the sunspot to that in the
quiet Sun, we have a closed system of three equations with three
variables: absorption, emissivity reduction, and local suppression.
By computing ratios of measured values in the sunspots to that
in the quiet Sun, we have avoided normalizing our computations, thus
making our derivations of those coefficients more accurate. The
cross-covariances of ingoing, outgoing, and two-skip waves for the
quiet Sun and the sunspot are given in Equations
(\ref{mark1})--(\ref{mark2}). The solution of the system is given in
equations (\ref{absorption})--(\ref{suppression}). We should note
here that although the waves in Figure \ref{energy_budget} are
associated with a particular travel distance, the coefficients of
absorption, emissivity reduction, and local suppression, as defined
in Equations (\ref{absorption})--(\ref{suppression}) are functions of the
travel distance $r$. We have

\begin{equation}
F_{\mathrm{out}}^{q}=\sqrt{1-d}I \label{mark1}
\end{equation}

\begin{equation}
F_{\mathrm{in}}^{q}=\sqrt{1-d}I
\end{equation}

\begin{equation}
F_{\mathrm{2skip}}^{q}=(1-d)I
\end{equation}

\begin{equation}
F_{\mathrm{in}}^{s}=\sqrt{(1-s)(1-a)(1-d)}I
\end{equation}

\begin{equation}
F_{\mathrm{out}}^{s}=\sqrt{(1-d)(1-a)(1-s)}[(1-a)(1-d)+(1-r)d]I
\end{equation}

\begin{equation}
F_{\mathrm{2skip}}^{s}=(1-d)(1-a)I \label{mark2}
\end{equation}

\begin{equation}
a=1-\frac{F_{\mathrm{2skip}}^{s}}{F_{\mathrm{2skip}}^{q}} \label{absorption}
\end{equation}

\begin{equation}
r=1-\frac{1}{d}\cdot\left[{\frac{F_{\mathrm{out}}^{s}}{F_{\mathrm{out}}^{q}}\frac{F_{\mathrm{in}}^{q}}{F_{\mathrm{in}}^{s}}-\frac{F_{\mathrm{2skip}}^{s}}{F_{\mathrm{2skip}}^{q}}(1-d)}\right]
\label{reduction}
\end{equation}

\begin{equation}
s=1-\left(\frac{F_{\mathrm{in}}^{s}}{F_{\mathrm{in}}^{q}}\right)^2\frac{F_{\mathrm{2skip}}^{q}}{F_{\mathrm{2skip}}^{s}}
\label{suppression}
\end{equation}

\par A similar procedure is followed for one-skip waves. The
ray path of these waves is illustrated in the bottom panel of
Figure \ref{energy_budget}. The main difference of the one-skip
waves from the two-skip waves is that the one-skip waves encounter
the sunspot in deeper layers and do not have upper turning points
located inside the magnetized photosphere, thus they can probe the
absorption efficiency of the sunspot at these depths. It is natural
to define a new absorption coefficient for the one-skip waves as
$a_{d}=1-(F_{\mathrm{1skip}}^{s}/F_{\mathrm{1skip}}^{q})^{2}$. However, using the same
definition for the absorption coefficient as the one used for the
two-skip waves allows for a direct comparison of the results
obtained with these two methods. The absorption as measured by the
one-skip waves is called hereafter ``deep absorption" to be
distinguished from the ``surface absorption" measured by the
two-skip waves and defined in Equation (\ref{absorption}). The ``deep
absorption" is defined similarly as

\begin{equation}
a_{d}=1-\frac{F_{\mathrm{1skip}}^{s}}{F_{\mathrm{1skip}}^{q}}
\label{surface_absorption}
\end{equation}

\par The propagation of acoustic waves in the Sun is affected by
dispersion. Dispersion increases the width and hence decreases the
amplitude of the cross-covariance. However, if dissipation and
absorption are ignored, the product of the square of the amplitude
and the width of the cross-covariance is constant (Chou and
Ladenkov, 2007; Burtseva \emph{et al.}, 2007). In order to correct
the effect of dispersion, the width-corrected cross-covariance
$\tilde{F}_{ab}$ is defined as
\begin{equation}
\tilde{F}_{ab}=F_{ab}(w_{ab})^{1/2} \label{dispersion}
\end{equation}
where $w_{ab}$ is the ratio of the width of the cross-covariance at
point $b$ to that at point $a$. All of the cross-covariances presented
in this paper are width corrected.

\section{Data Analysis and Results}
\par Doppler observations from MDI onboard the
\emph{Solar and Heliospheric Observatory} (Scherrer \emph{et al.},
1995) are used in this work. The study of ingoing, outgoing, one-
and two-skip waves both for the active regions and the quiet Sun
utilizes 31 datasets, selected from 1996 to 2001. Each dataset is
512 minutes long, tracked with a Carrington rotation rate and
remapped to Postel's coordinates centered at the main active region,
with a spatial resolution of $0.12^\circ$ pixel$^{-1}$ and a size of
$256 \times 256$ pixels. Each dataset is then filtered in the
Fourier domain to remove solar convection and \textit{f} modes. We do not
filter out signals above the cut-off frequency since most of
acoustic power is concentrated below the cut-off limit and our
experiments also show that with and without filtering those signals,
our results do not change.
\par
\begin{figure}[h]
\centerline{\includegraphics[width=0.5\textwidth,clip=]{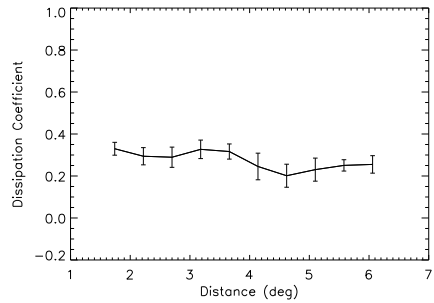}}
\caption{Dissipation coefficient in the quiet Sun as a function of
acoustic wave-travel distance.} \label{dissipation_figure}
\end{figure}

 The computational procedure starts with the determination of
dissipation in the quiet Sun that appears in Equation
(\ref{reduction}). We select 900 pixels from each dataset to compute
center-to-annulus cross-covariances with the time--distance
helioseismology technique (Duvall \emph{et al.}, 1993; Kosovichev,
Duvall, and Scherrer, 2000). Two concentric quadrants are selected
around a central pixel so that the radius of the larger quadrant is
five times larger than that of the smaller one. The signal inside
the quadrants is averaged, and the cross-covariance between the
signal of the central pixel and the averaged signal of each quadrant
is computed for both positive and negative travel-time lags. Each
cross-covariance is then multiplied by the length of the
corresponding quadrant because, in the absence of a direction filter,
the acoustic energy, propagating from the central pixel to the
quadrant, is uniformly distributed over it. The same procedure is
repeated for a range of radii of $11-45$ pixels and all of the
cross-covariances for the same distances and for both time lags,
obtained from different central pixels and different datasets, are
combined to increase the signal-to-noise ratio. For larger travel
distances of $46-54$ pixels, the use of five skips is not possible
due to the limited size of the dataset and thus only four skips were
used. Equation (\ref{dissipation}) is modified for this case to
$d(r)=1-(F_{4}/F_{1})^{2/3}$. The one- and the five-skip (or
four-skip) signals are fitted with a Gabor wavelet function
(Kosovichev and Duvall, 1996) and the amplitudes and widths of the
two cross-covariances are obtained. The dissipation is calculated
using Equations (\ref{dissipation}) and (\ref{dispersion}) and the
result is presented in Figure \ref{dissipation_figure}.

\par We should point out that an accurate measurement of dissipation
is crucial for the work that follows since the dissipation coefficient
appears in Equation (\ref{reduction}) that is used to determine the
emissivity reduction. The approach followed here is based on
measurements of an amplitude ratio of five-skip (or four-skip)
signals to one-skip signals. In fact any number of skips can be used
to determine the dissipation coefficient. Therefore, the ratio of
five-skip signal-to-one-skip signal in Equation (\ref{dissipation})
can be a ratio of two skips to one skip or three skips to two skips
\textit{etc}. Since after each skip the acoustic energy is reduced by a
factor of $(1-d(r))$, the larger the difference of skips in the
ratio of Equation (\ref{dissipation}), the larger the dissipated
energy. So using more skips makes our measurements more accurate.
The use of more than five-skip signals is not possible due to the
limited size of the dataset, so for the dissipation coefficient
presented in Figure 2, five skips are used for distances up to
$5.4^\circ$ and four skips for larger distances. The two methods, as
well as others with different combinations of skips, are consistent
with the only difference being that the signal-to-noise ratio is higher for
those methods where the dissipated energy is larger.

\par Next we determine the absorption, emissivity reduction, and
local suppression coefficients inside active regions. This study
includes $47$ active regions distributed in $31$ datasets. The
strength of the magnetic field in the magnetograms, which were obtained
at the same time with same resolution as the Dopplergrams, is used as the
criterion to select the pixels inside the active region in the
corresponding Dopplergrams. Setting the threshold at $1100$ G and
$500$ G, we select approximately the pixels inside the umbra and
both the umbra and penumbra (sunspot) respectively. Each one of
these pixels is used to compute center-to-annulus cross-covariances.
An annulus around the central pixel is selected and the signal
inside this annulus is averaged. The cross-covariance between the
central pixel and the averaged signal inside the annulus is computed
for both positive and negative travel-time lags. The positive
travel-time lag corresponds to outgoing waves and the negative
travel-time lag corresponds to ingoing waves. The same procedure is
repeated for a range of radii of $11-54$ pixels and all of the
cross-covariances for the same distances, obtained from different
central pixels and different datasets, are combined to increase the
signal-to-noise ratio. The signals corresponding to outgoing and
ingoing waves are fitted with a Gabor wavelet function and the
amplitudes and widths of the two signals are obtained.

\begin{figure}[h]    
   \centerline{\hspace*{0.015\textwidth}
               \includegraphics[width=0.815\textwidth,clip=]{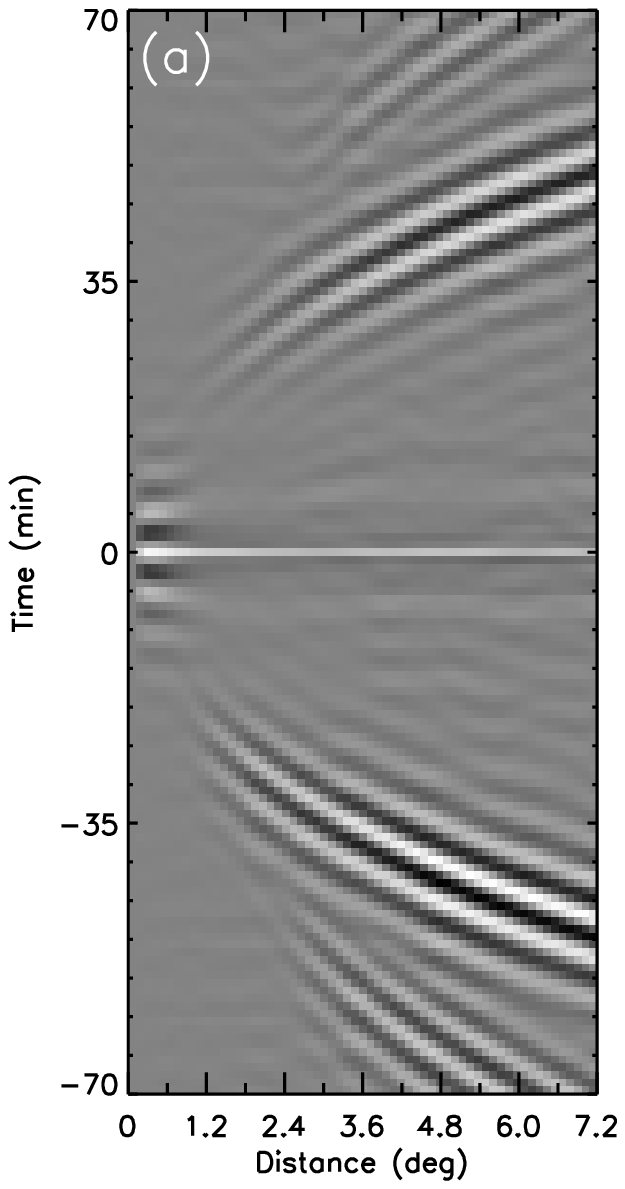}
               \hspace*{-0.33\textwidth}
               \includegraphics[width=0.815\textwidth,clip=]{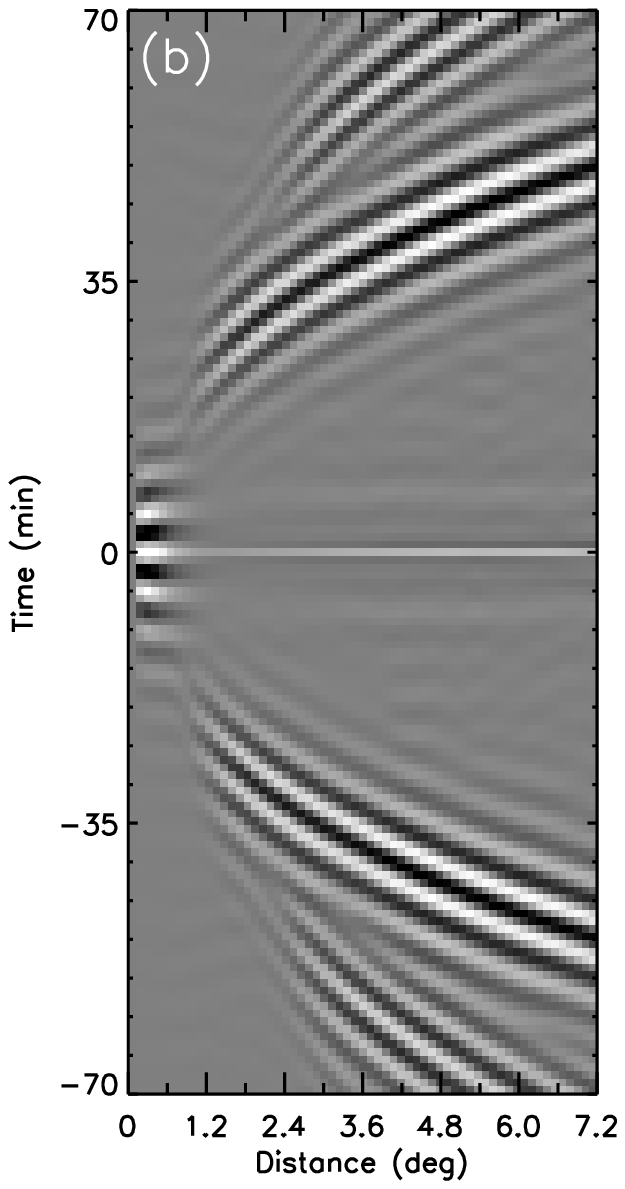}
              }
     \vspace{0.\textwidth}   
   \centerline{\hspace*{0.015\textwidth}
               \includegraphics[width=0.515\textwidth,clip=]{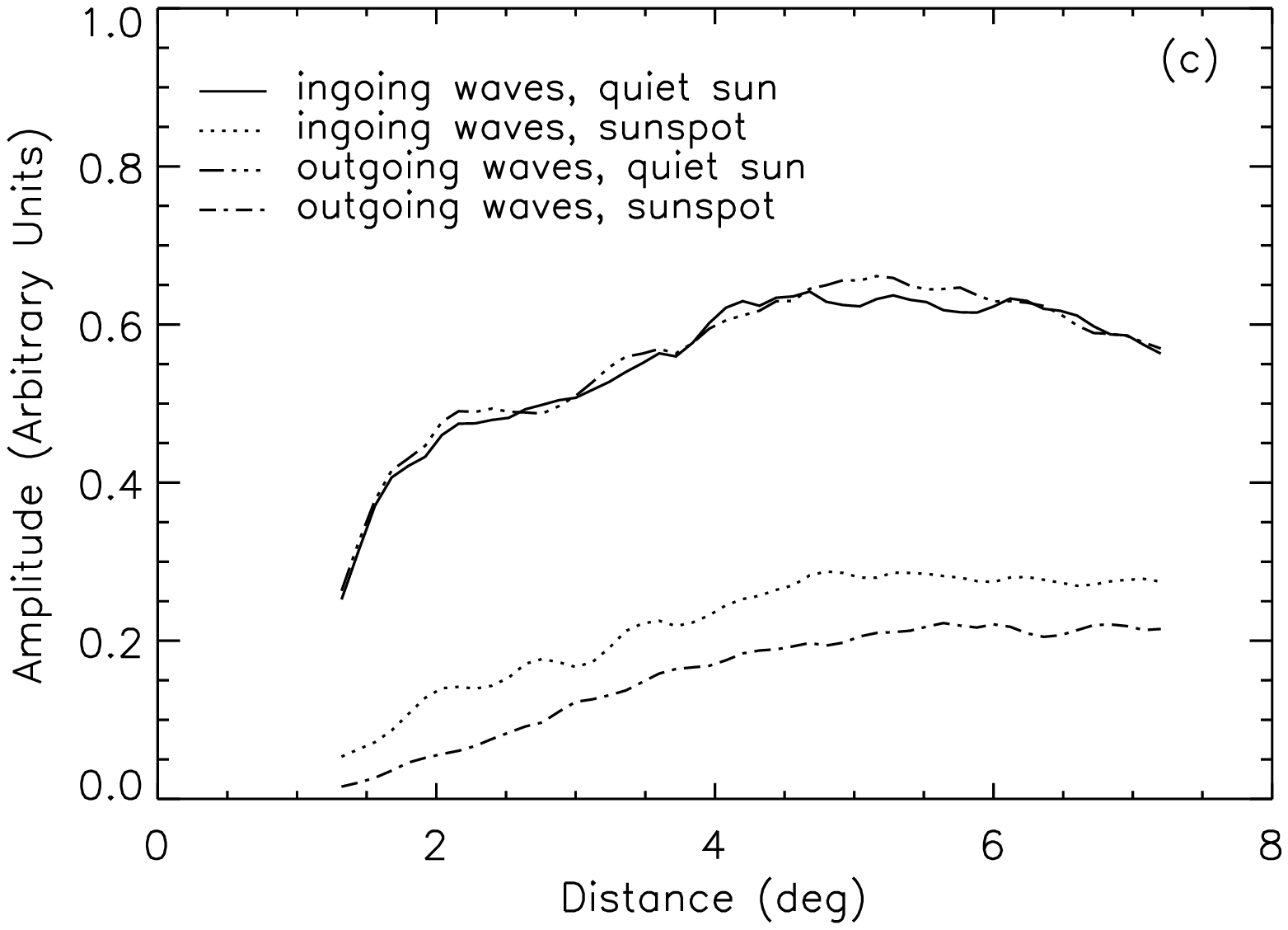}
              }
     \vspace{0.\textwidth}   

     \caption{Time--distance diagrams of ingoing and outgoing waves for
     (a) sunspots and (b) quiet Sun. The positive travel--time lag corresponds
     to outgoing waves and the negative to ingoing waves. The amplitude fittings of the
     cross-covariance function are shown in panel (c).
        }
   \label{1skip_figure}
   \end{figure}

\par
\begin{figure}[ht]    
   \centerline{\hspace*{0.015\textwidth}
               \includegraphics[width=0.815\textwidth,clip=]{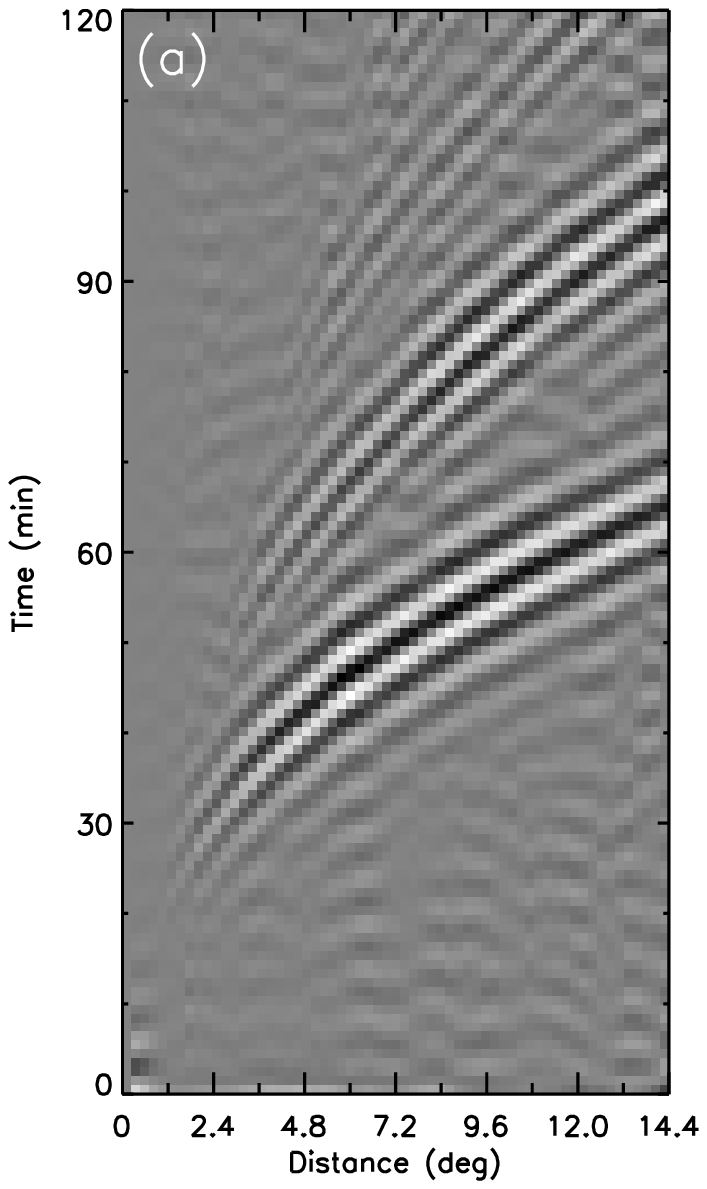}
               \hspace*{-0.33\textwidth}
               \includegraphics[width=0.815\textwidth,clip=]{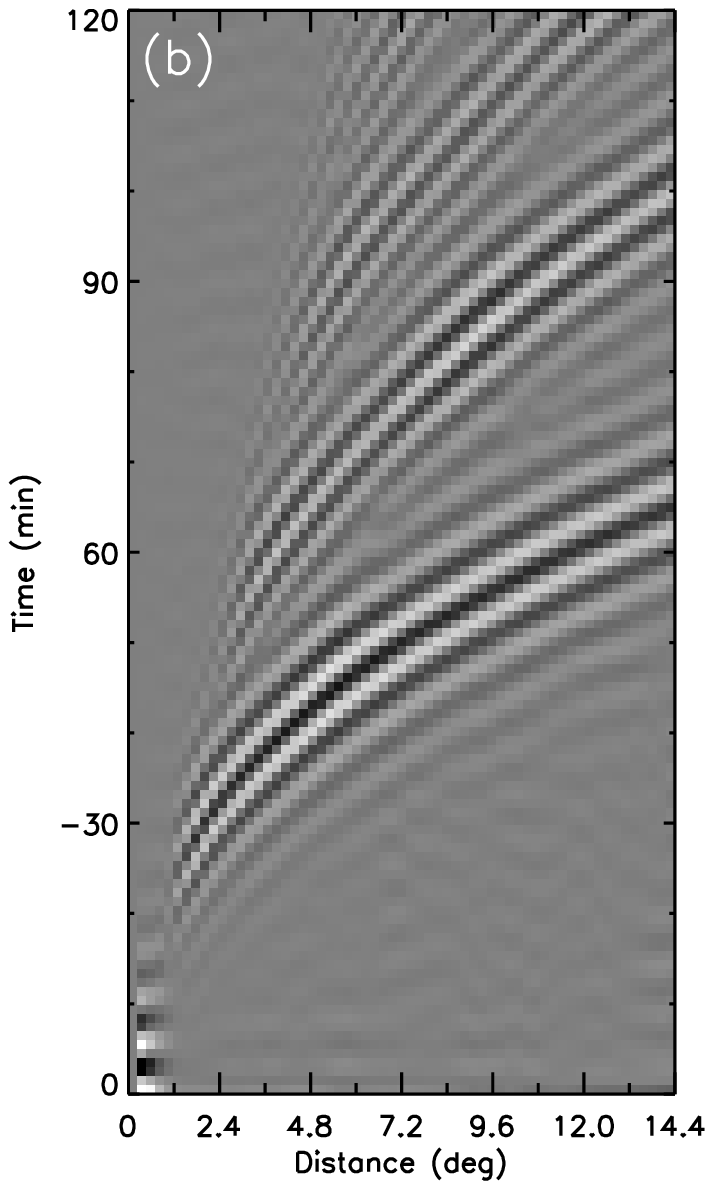}
              }
     \vspace{0.\textwidth}    
   \centerline{\hspace*{0.015\textwidth}
               \includegraphics[width=0.515\textwidth,clip=]{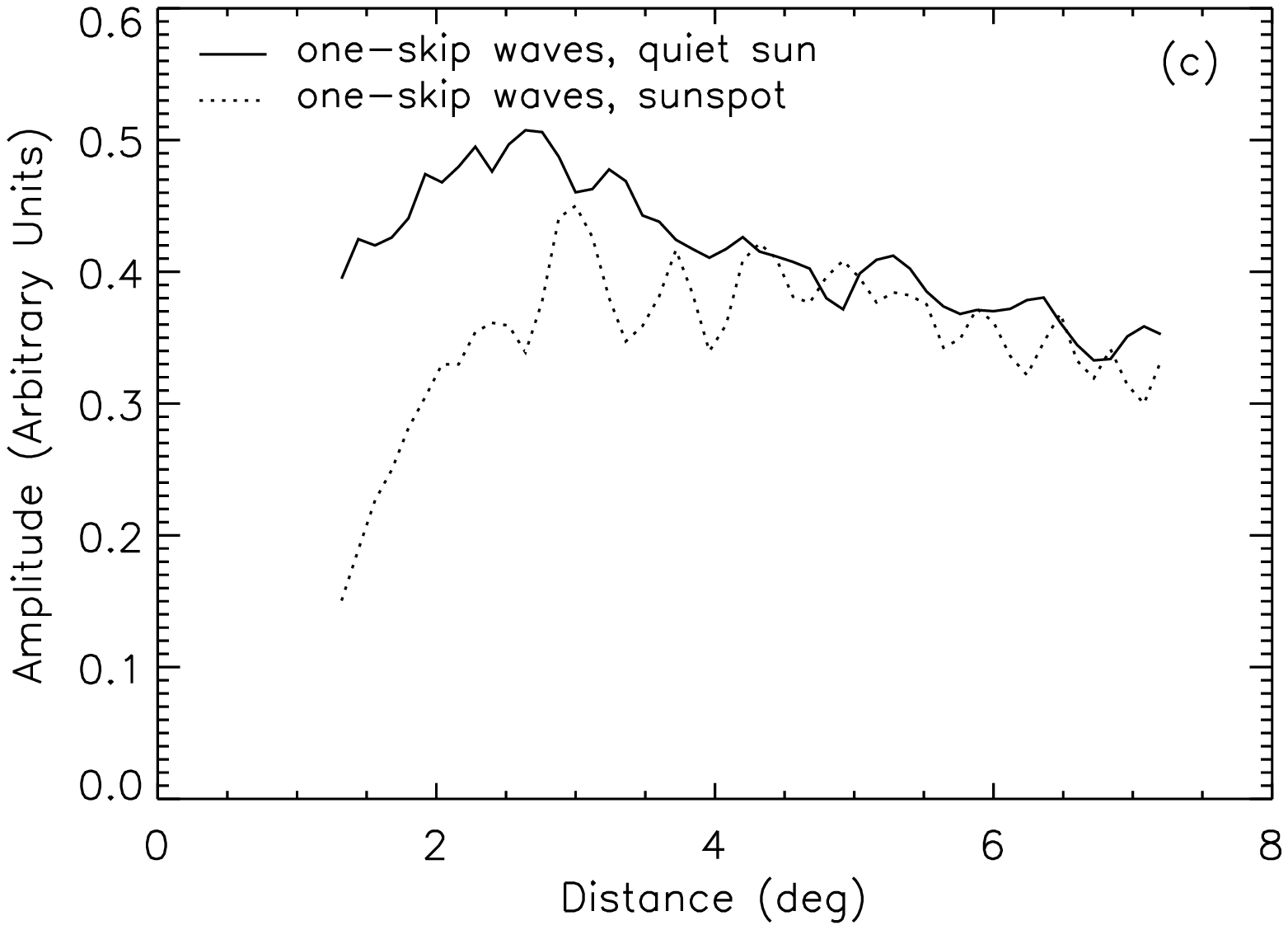}
               \hspace*{-0.03\textwidth}
               \includegraphics[width=0.515\textwidth,clip=]{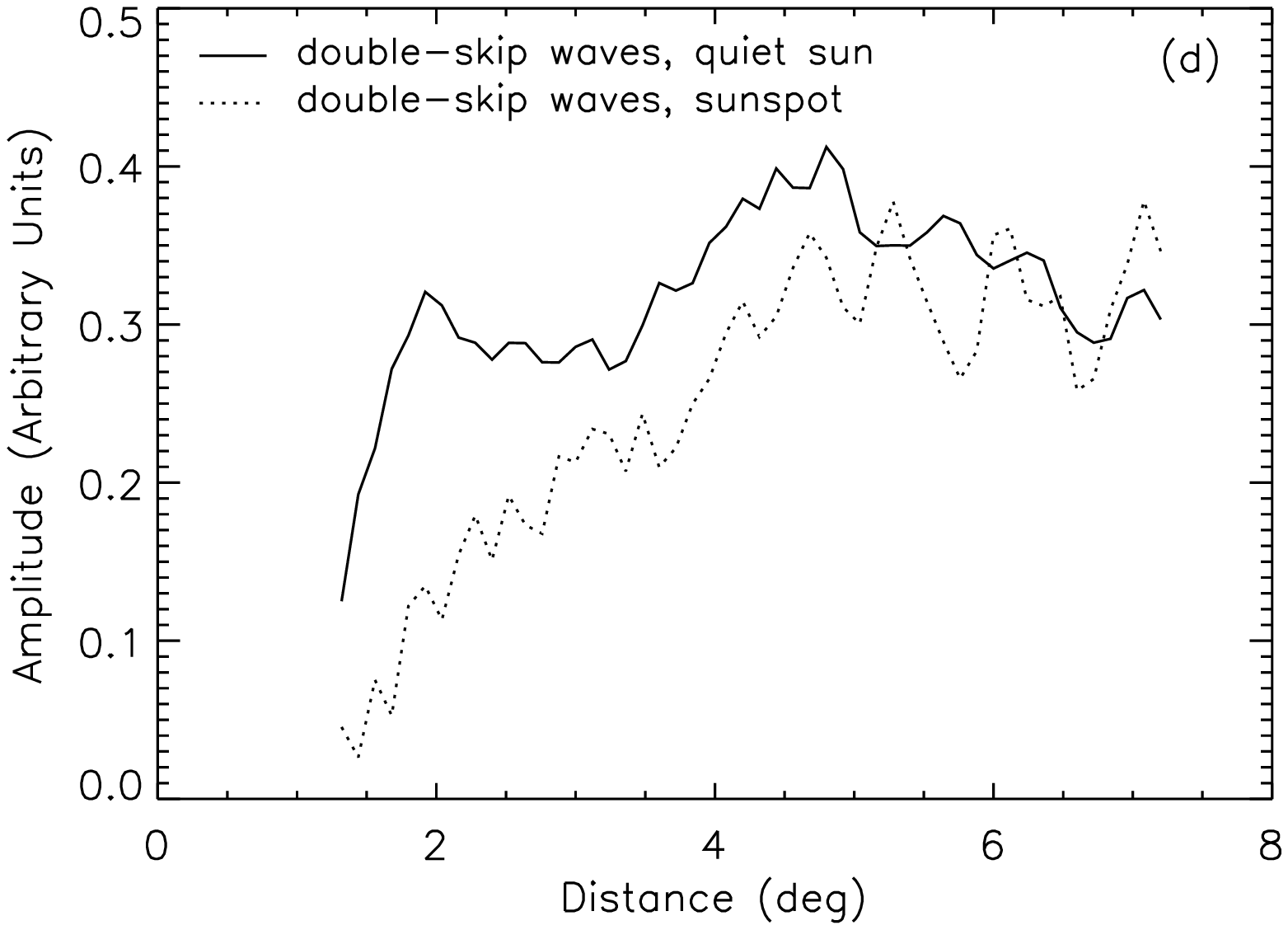}
              }
     \vspace{-0.\textwidth}    

     \caption{Time--distance diagrams of one- and two-skip measurements for
     (a) sunspots and (b) quiet Sun. Amplitude fitting of the
     cross-covariance function for (c) one-skip waves and (d) two-skip waves.
        }
   \label{2skip_figure}
   \end{figure}

\begin{figure}[ht]    
   \centerline{\hspace*{0.015\textwidth}
               \includegraphics[width=0.515\textwidth,clip=]{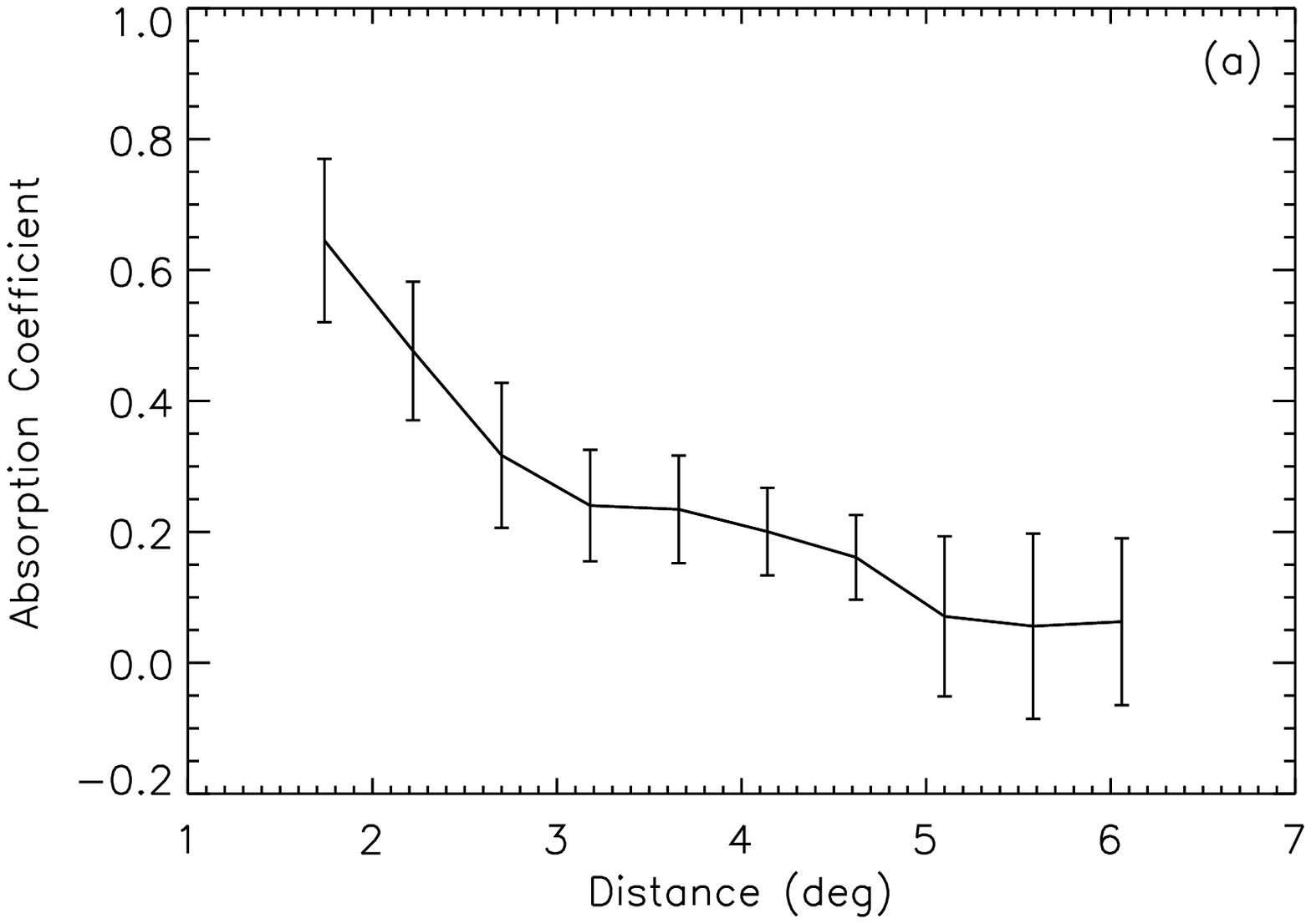}
               \hspace*{-0.03\textwidth}
               \includegraphics[width=0.515\textwidth,clip=]{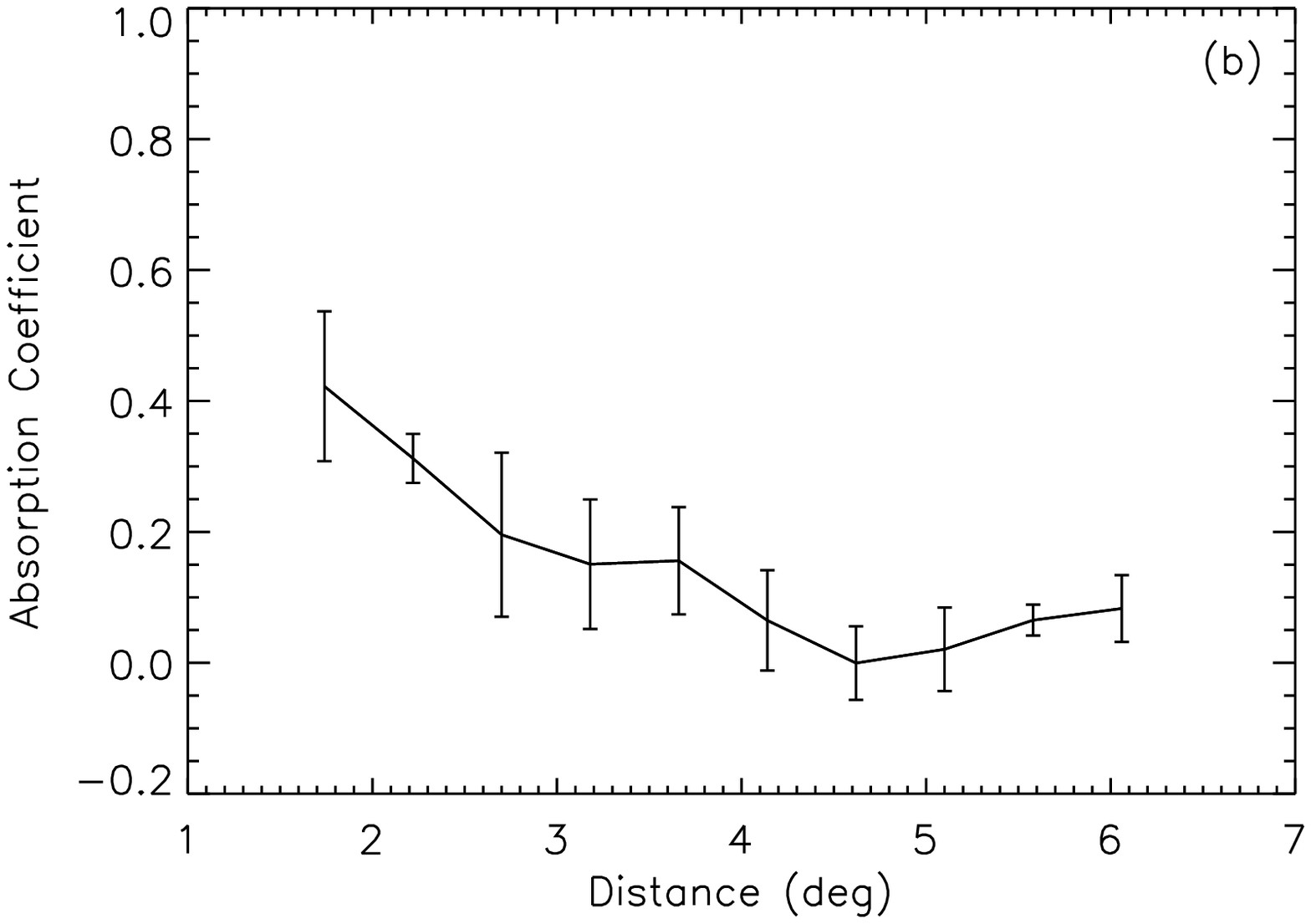}
              }
     \vspace{0.0\textwidth}   
   \centerline{\hspace*{0.015\textwidth}
               \includegraphics[width=0.515\textwidth,clip=]{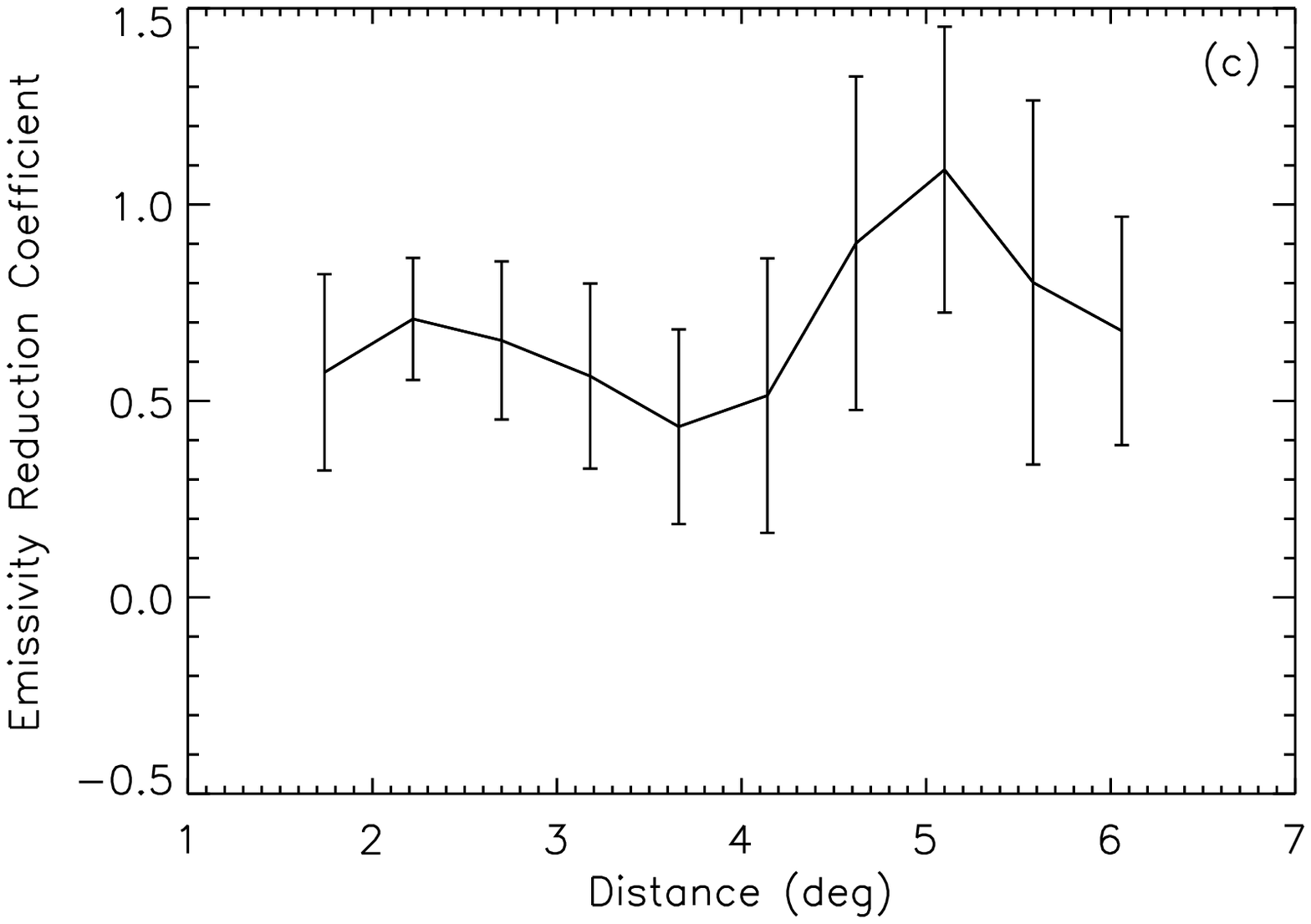}
               \hspace*{-0.03\textwidth}
               \includegraphics[width=0.515\textwidth,clip=]{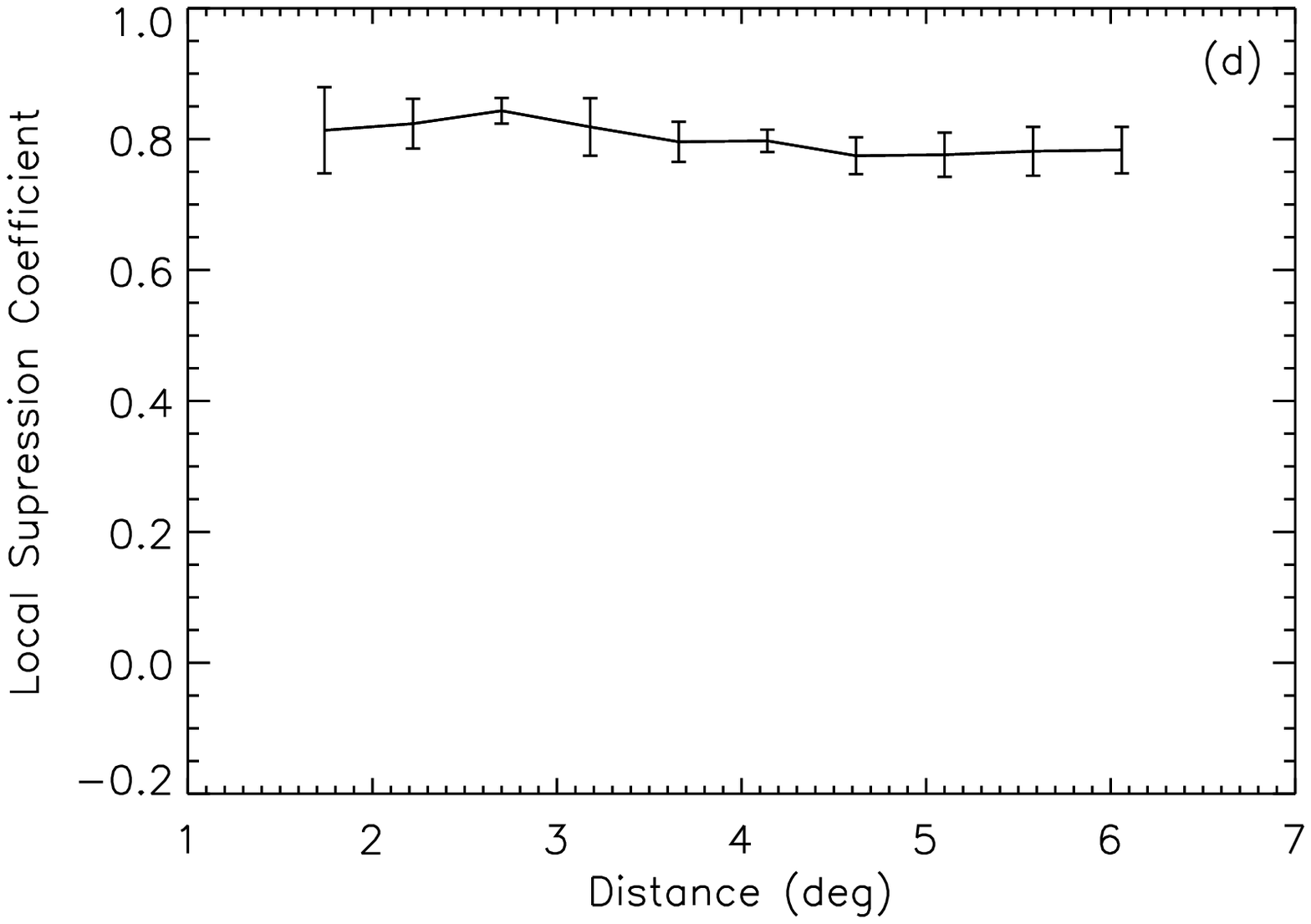}
              }
     \vspace{0.0\textwidth}    

     \caption{Coefficients of (a) surface absorption, (b) deep
     absorption, (c) emissivity reduction, and (d)
     local suppression obtained from umbral
     measurements (threshold of $1100$ G).
        }
   \label{results_1100}
   \end{figure}

\begin{figure}[h]    
   \centerline{\hspace*{0.015\textwidth}
               \includegraphics[width=0.515\textwidth,clip=]{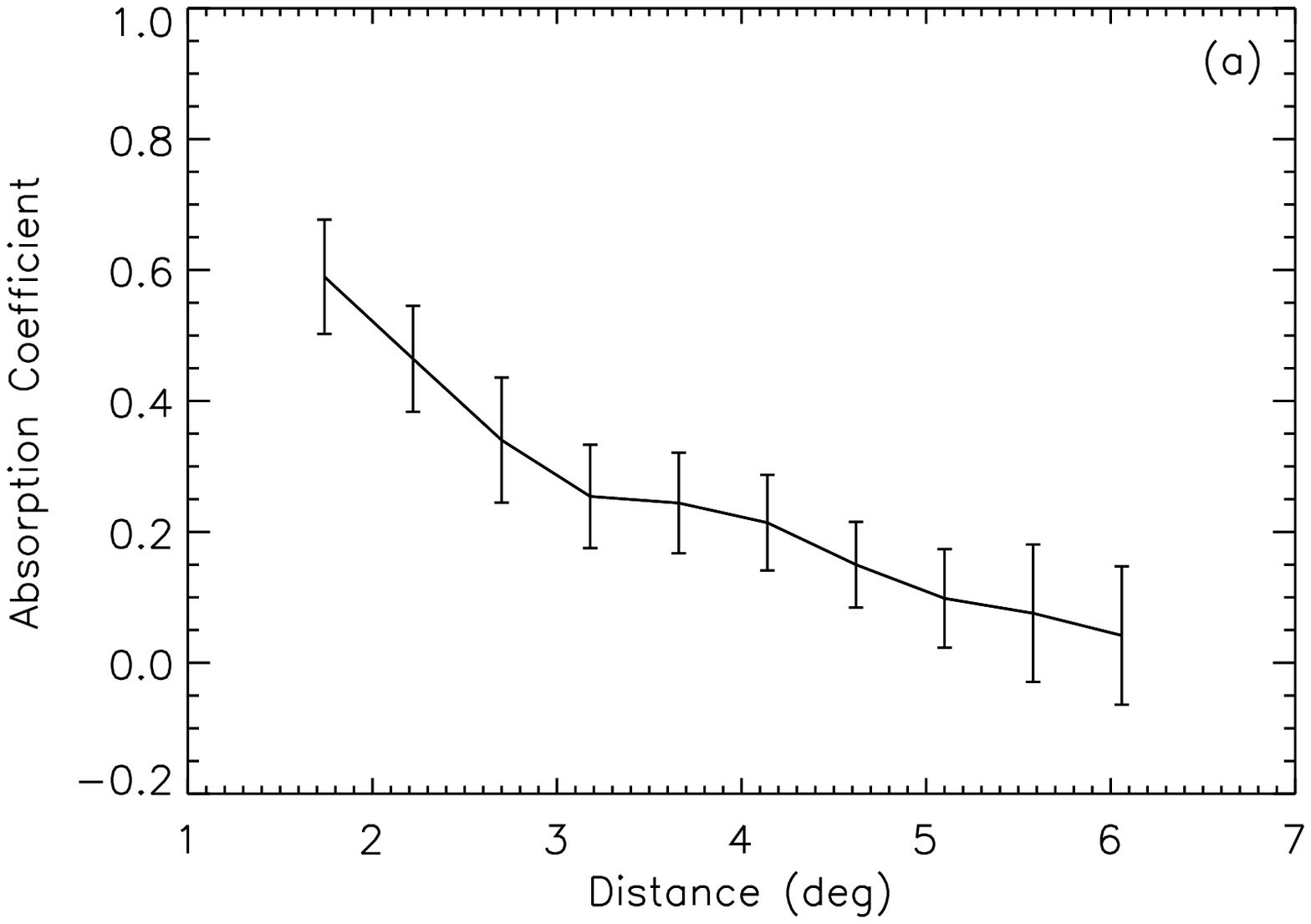}
               \hspace*{-0.03\textwidth}
               \includegraphics[width=0.515\textwidth,clip=]{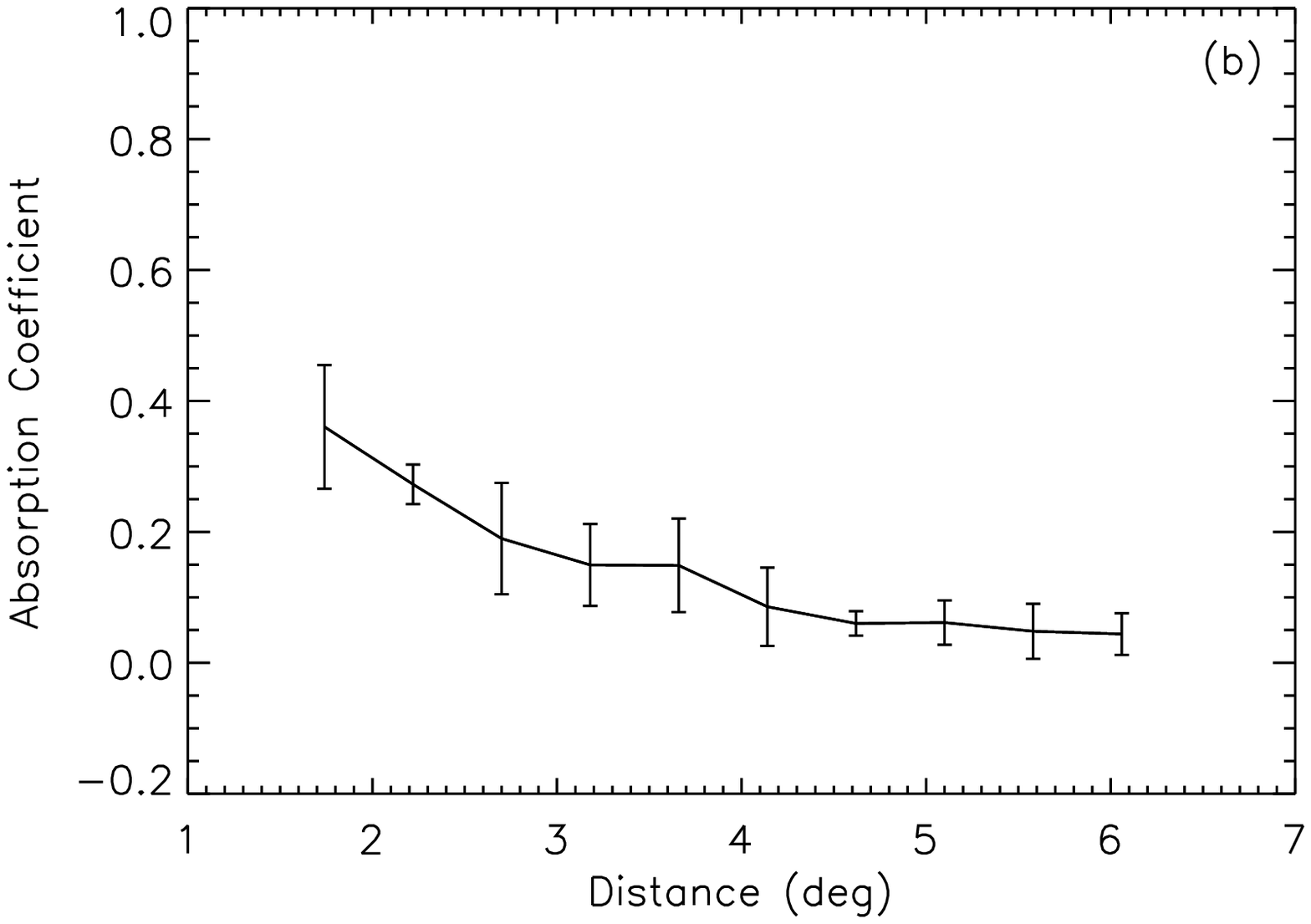}
              }
     \vspace{0.\textwidth}   
   \centerline{\hspace*{0.015\textwidth}
               \includegraphics[width=0.515\textwidth,clip=]{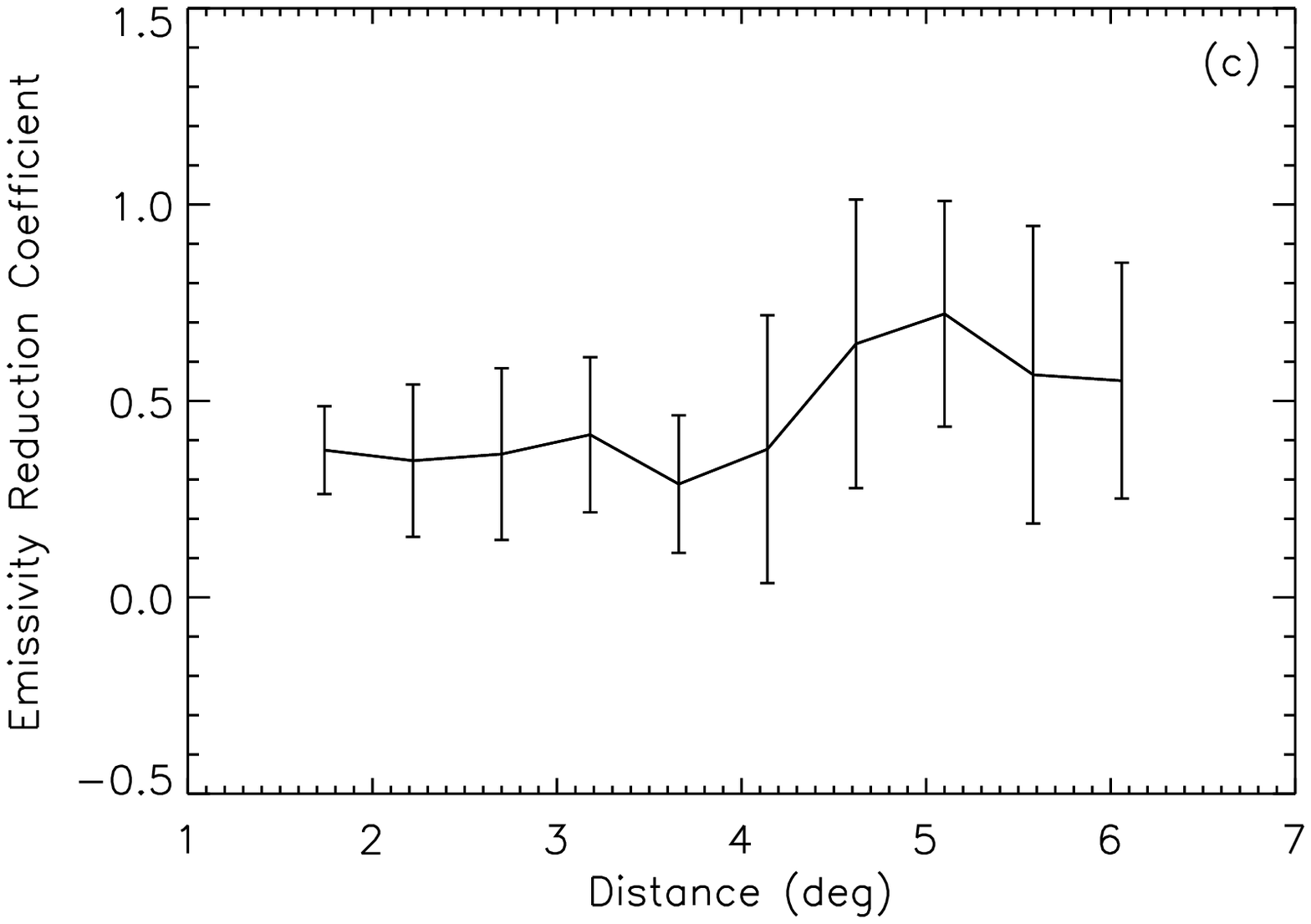}
               \hspace*{-0.03\textwidth}
               \includegraphics[width=0.515\textwidth,clip=]{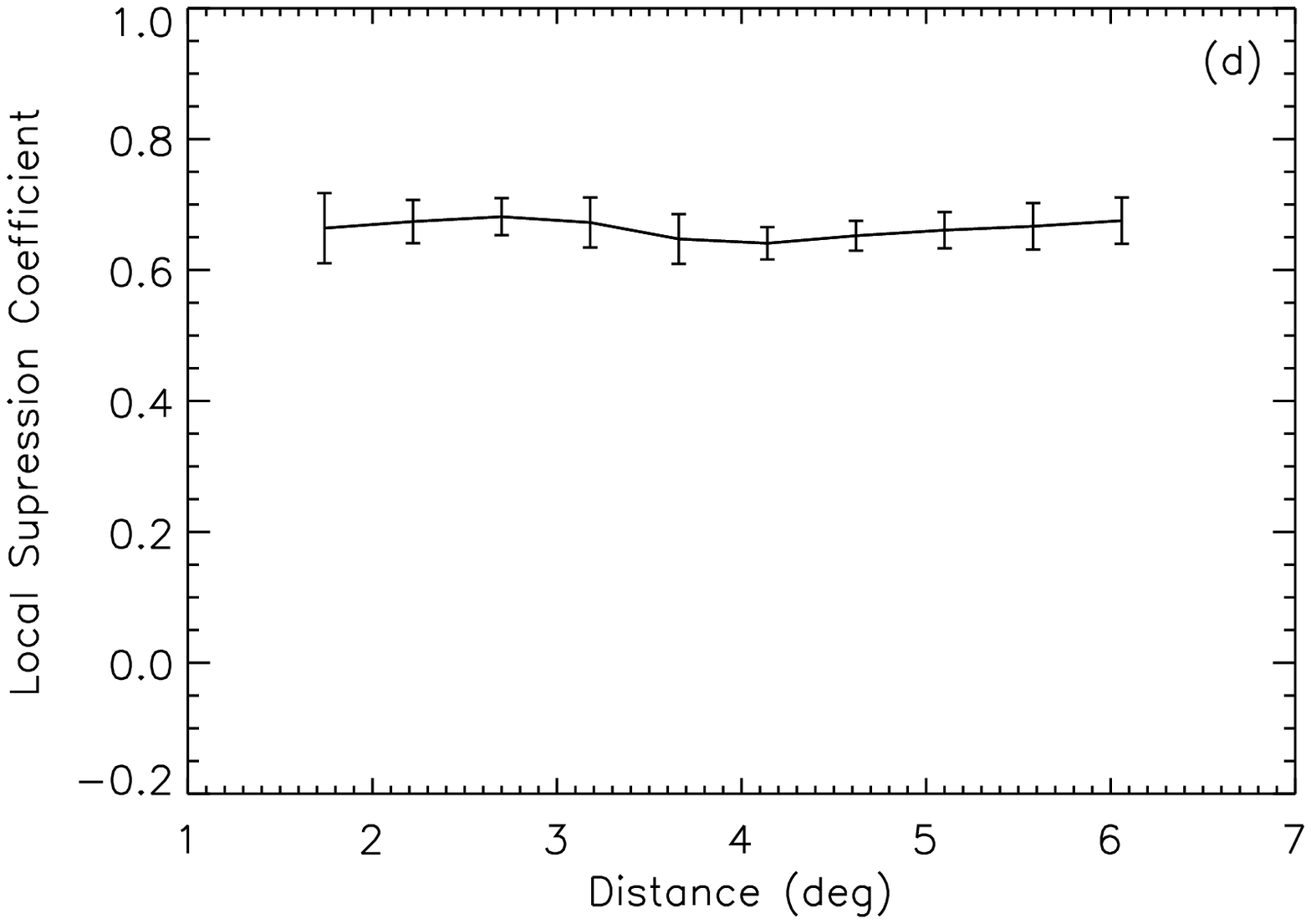}
              }
     \vspace{0.0\textwidth}   

     \caption{Coefficients of (a) surface absorption, (b) deep
     absorption, (c) emissivity reduction,
     and (d) local suppression obtained from sunspot
     measurements (threshold of $500$ G).
        }
   \label{results_500}
   \end{figure}

 For the one- and two-skip waves, the
annulus around the central pixel is divided into four quadrants and
the signal inside each one of them is averaged. The cross-covariance
is computed between the two pairs of diametrically opposite
quadrants for both positive and negative travel-time lags. The
process is repeated again for the same range of radii and all the
cross-covariances for the same distances and for both time lags,
obtained from different central pixels and different datasets, are
combined. The two-skip signal is again fitted with a Gabor wavelet
function and the amplitude and width of the signal are obtained. The
exactly same procedures are carried out for the quiet Sun to obtain
cross-covariances as references. The time--distance diagrams for the
sunspot and the quiet Sun as well as the amplitude fittings of the
ingoing and outgoing waves are shown in Figure \ref{1skip_figure}.
The corresponding time--distance diagrams and amplitude fittings for
the one- and two-skip waves are shown in Figure \ref{2skip_figure}.
The three coefficients of absorption, emissivity reduction, and local
suppression are computed using Equations
(\ref{absorption})--(\ref{suppression}) and the results are presented
in Figures \ref{results_1100} and \ref{results_500}. The error bar
corresponds to the standard deviation of those measurements from
which mean values are obtained.

\par In the model of Figure 1, the beginning point is assumed to have
the power of the quiet Sun. If that is not the case, an error will arise in the
computation of the cross-covariances and the coefficients of absorption,
emissivity reduction, and local suppression (Chou \emph{et al.}, 2009d). In
order to make our measurements as accurate as possible, the computation of
cross-covariances includes only those pixels for which the magnetic field at
the beginning point of Figure 1 is lower than an arbitrary threshold, which is set to 40 G.
For the selection of pixels in the quiet Sun, the closest magnetic region
to the quiet pixel is at least nine pixels away from it. The magnetic region
is defined as a region where the magnetic field is higher than 500 G.
For some large active regions, the annuli of short distances are
located completely inside the sunspot and thus no pixel from those annuli can be used
for the computation of center-to-annulus cross-covariances. In this case we use the average cross-covariance
with same annulus distance from the central pixel, as calculated from all other central
pixels of the same dataset. The algorithm described above
makes sure that the computation of cross-covariances for the sunspot and the
quiet Sun uses the exactly same number of annuli. However, it is possible that at
short distances a significant number of pixels in the annuli are excluded from the
computation of cross-covariances and therefore the accuracy of the
measurements can be affected.

\par The measurements of ingoing and outgoing waves are less noisy
than those of one- and two-skip waves. This can be seen either from
the time--distance diagrams in Figures \ref{1skip_figure} and
\ref{2skip_figure} or from the amplitude fittings of those diagrams.
However, in both cases the amplitude and width of the
cross-covariance function can easily be obtained by our method. For
the quiet Sun, ingoing and outgoing waves have approximately equal
amplitudes. On the contrary, for sunspot measurements the amplitude of
outgoing waves is reduced compared to that of ingoing waves. The
one- and two-skip wave amplitudes show a different trend. At short
travel distances, the wave amplitude in the quiet Sun is much greater
than that in sunspot but at large travel distances the two
amplitudes are comparable within the noise level of the
measurements.

\par The surface absorption coefficient, presented in
Figures \ref{results_1100}a and \ref{results_500}a, is a maximum at the
shortest travel distances and smoothly drops to zero at distance of
about $6^\circ$. This picture is consistent with previous work (for
example Braun, Duvall, and LaBonte, 1987, 1988; Bogdan \emph{et
al.}, 1993; Chen \emph{et al.}, 1996). However, a quantitative
comparison is not possible, first due to the different definitions of
absorption and second because our definition separates the effects
of absorption and emissivity reduction. The deep absorption follows
the same trend but compared to the surface absorption is about
$40\%$ smaller at small travel distances where both of them are
greater than zero. At large travel distances they both drop to zero
and become comparable. The emissivity reduction coefficient is most
noisy due to the several terms involved in Equation
(\ref{reduction}). Inside umbrae emissivity reduction is rather high
ranging from $0.44$ to $1.00$ with a mean value of $0.70$, while in
the whole sunspot the corresponding range is $0.29$ to $0.72$
averaging $0.47$. The local suppression coefficient is very
constant within the range of travel distances used in this work. Its
mean values for umbral and penumbral measurements are $0.80$ and
$0.665$ respectively.

\par
\begin{figure}[ht]    
   \centerline{\hspace*{0.015\textwidth}
               \includegraphics[width=0.515\textwidth,clip=]{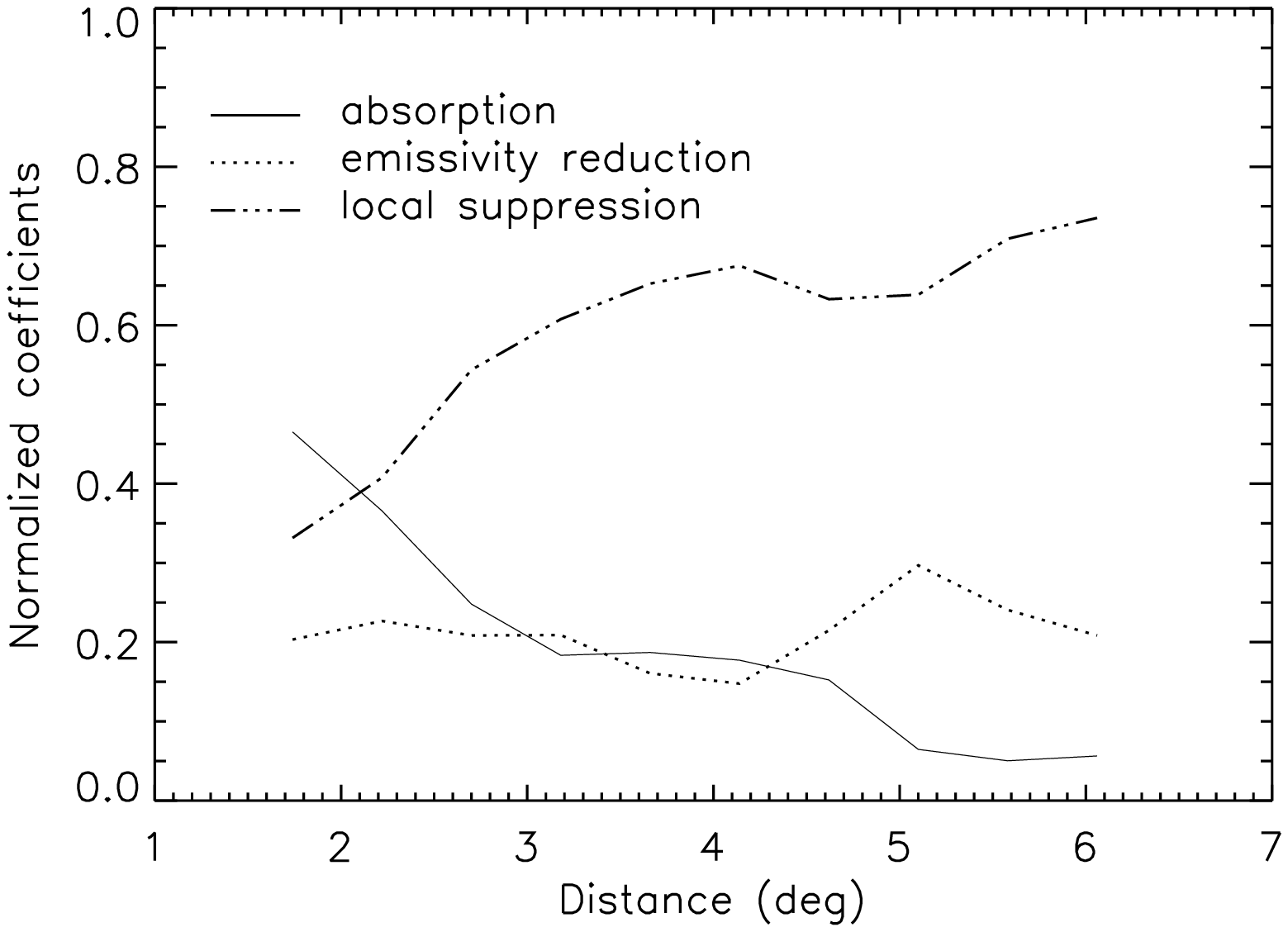}
               \hspace*{-0.03\textwidth}
               \includegraphics[width=0.515\textwidth,clip=]{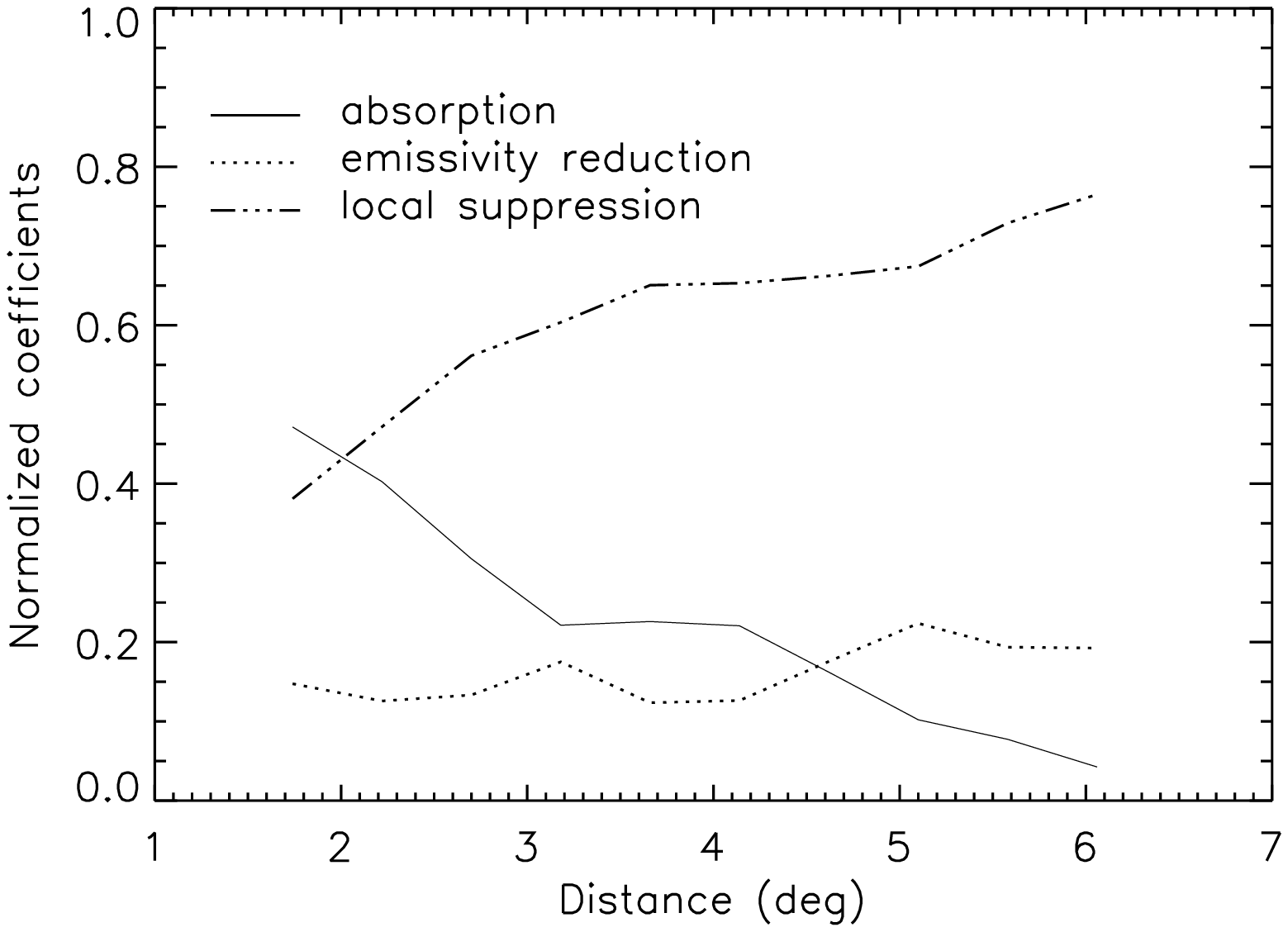}
              }
              \vspace{0.0\textwidth}   

     \caption{Normalized coefficients as functions of the travel
     distance for the umbral measurements (left) and the sunspot
     measurements (right).
        }
   \label{norm_results}
   \end{figure}

 All three coefficients of absorption, emissivity reduction, and local
suppression, as defined in this work, contribute to the power
deficit inside sunspots. However, it is not clear, from this
definition, what is the exact contribution of each coefficient to the total
power deficit in sunspots. It is more appropriate to define three
normalized coefficients so that each is equal to the fraction of
the contribution of the corresponding mechanism to the total power
deficit inside the sunspot. According to the energy budget of the
acoustic waves in the sunspot, the normalized coefficients are
\begin{equation}
a_{*}=\frac{a(1-d)}{1-P} \label{norm_absorption}
\end{equation}
\begin{equation}
r_{*}=\frac{rd}{1-P} \label{norm_reduction}
\end{equation}
\begin{equation}
s_{*}=\frac{s[(1-a)(1-d)+(1-r)d]}{1-P} \label{norm_suppression}
\end{equation}
where $P=(1-s)(1-a)(1-d)+(1-s)(1-r)d$ is the total acoustic power
inside the sunspot. It is easily shown that $a_{*}+r_{*}+s_{*}=1$.
The normalized coefficients are plotted as functions of travel
distance in Figure \ref{norm_results}.
\par The definition of normalized coefficients allows a direct
comparison of the results obtained in this work with results from
previous works. Chou \emph{et al.} (2009c), using the same
definition, found that the fractional contribution of each mechanism
to the acoustic power deficit in the umbra of the sunspot for a
specific wave packet that corresponds to travel distance of
$3.5^\circ$ is $a=0.233$, $s=0.685$ and $r=0.082$. For the same
travel distance, we found $a=0.187$, $s=0.652$ and $r=0.161$. The two
methods are consistent in local suppression but not in absorption
and emissivity reduction. The absorption in Chou \emph{et al.} is
larger compared to our work while the emissivity reduction is
smaller. However, the results obtained in this work are based on
averaged measurements over $46$ sunspots while the results presented
by Chou \emph{et al.} are based on a single sunspot. Not only are the
two samples different but it is also possible that there are
significant variations among the $47$ sunspots analyzed in this
study. Especially the coefficients of absorption and emissivity
reduction may have strong dependence on the size of the sunspot and
the strength of the magnetic field. It would be interesting to use the sample
of $47$ sunspots to study this dependence.
Our method does not make use of phase-speed and direction filters though, and consequently,
time--distance measurements with individual sunspots are very hard, if not impossible.
In practice, the low number of total pixels increases the noise level in the
cross-covariance function for the one- and especially for the two-skip waves,
making the measurement of the amplitude very hard.

\section{Summary}
We apply a new method to measure the coefficients of absorption,
emissivity reduction, and local suppression inside solar active
regions as well as the coefficient of dissipation in the quiet Sun.
This method utilizes measurements from many active regions to
increase the signal-to-noise ratio, and does not use signal filters
except filtering out solar convection and \textit{f}-modes. All four
coefficients are determined as functions of the travel distance.
Comparison with previous work for a specific travel distance shows
good agreement in the coefficient of local suppression but
discrepancies in coefficients of absorption and emissivity
reduction. It is not clear though if these discrepancies are caused
by the different method used in each case or by the different
samples of sunspots. In summary, our results show the following:
\begin{enumerate}
\renewcommand{\theenumi}{\roman{enumi}}
\renewcommand{\labelenumi}{\theenumi}

\item The absorption coefficient is the dominant
mechanism for the power deficit in sunspots at short distances but
it gradually drops to zero at travel distance of about $6^\circ$. \\
\item The measured deep absorption is about $60\%$ as high as the
measured surface absorption indicating that sunspots can actually absorb a large amount of
acoustic energy not only close to the surface
but also deep below the photosphere. \\
\item The emissivity reduction coefficient ranges between $0.44$ and
$1.00$ within the umbra with a mean value of $0.70$, and between
$0.29$ and $0.72$ within the sunspot with a mean value of $0.47$.
The fractional contribution to the acoustic power deficit in
sunspots is
$21.5\%$ for the umbra and $16.5\%$ for the sunspot. \\
\item The local suppression coefficient is remarkably constant as a
function of the travel distance with a value of $0.80$ for the
umbral measurements and $0.665$ for the sunspot measurements. Its
fractional contribution to the acoustic power deficit
increases smoothly as a function of the travel distance from $33\%$ to $75\%$
in the umbra and from $38\%$ to $79\%$ in the sunspot.
\end{enumerate}



%

%

%
%

%
%
%

\end{article}
\end{document}